\documentclass[a4paper,11pt]{article}
\usepackage{pos}
\usepackage{soul} %we can remove this when we are done
\usepackage{enumitem}

\title{SMEFT analysis of the electroweak sector: challenges beyond dimension 6 }
\author[a]{Raquel Gomez-Ambrosio}
\author*[b]{Giacomo Magni}

\affiliation[a]{Dipartimento di Fisica, Universita degli Studi di Milano Bicocca \\
and INFN Sezione di Milano Bicocca, Milan, Italy.}

\affiliation[b]{ Department of Physics and Astronomy, Vrije Universiteit Amsterdam, and Nikhef Theory Group,  Amsterdam, The Netherlands}

\emailAdd{raquel.gomezambrosio@unimib.it}
\emailAdd{g.magni@vu.nl }

\abstract{In these proceedings we present the latest developments in our effort to include vector boson scattering (VBS) measurements into global SMEFT fits of LHC data. We present some updates to our initial study of arXiv:2101.03180 as well as comment on a possible roadmap for the inclusion of higher orders beyond dimension 6 in the SMEFT  and on the interpretation of VBS data in other EFT frameworks.}

\FullConference{%
  *** The European Physical Society Conference on High Energy Physics (EPS-HEP2021), ***\\
  *** 26-30 July 2021 ***\\
  *** Online conference, jointly organized by Universität Hamburg and the research center DESY ***
}

\begin{document}
\maketitle

\section{Introduction}

In the past few years a significant amount of work has been done to extend the effort to make a global fit for the dimension 6 EFT basis to all the LHC dataset. After the initial attempts where only a few operators were considered and only Higgs channels included, we now have more sophisticated tools, able to fit a large number of operators simultaneously~\cite{Ellis:2020unq,Ethier:2021bye}.

\begin{figure}[!ht]
\begin{center}

    \includegraphics[width=.95\textwidth]{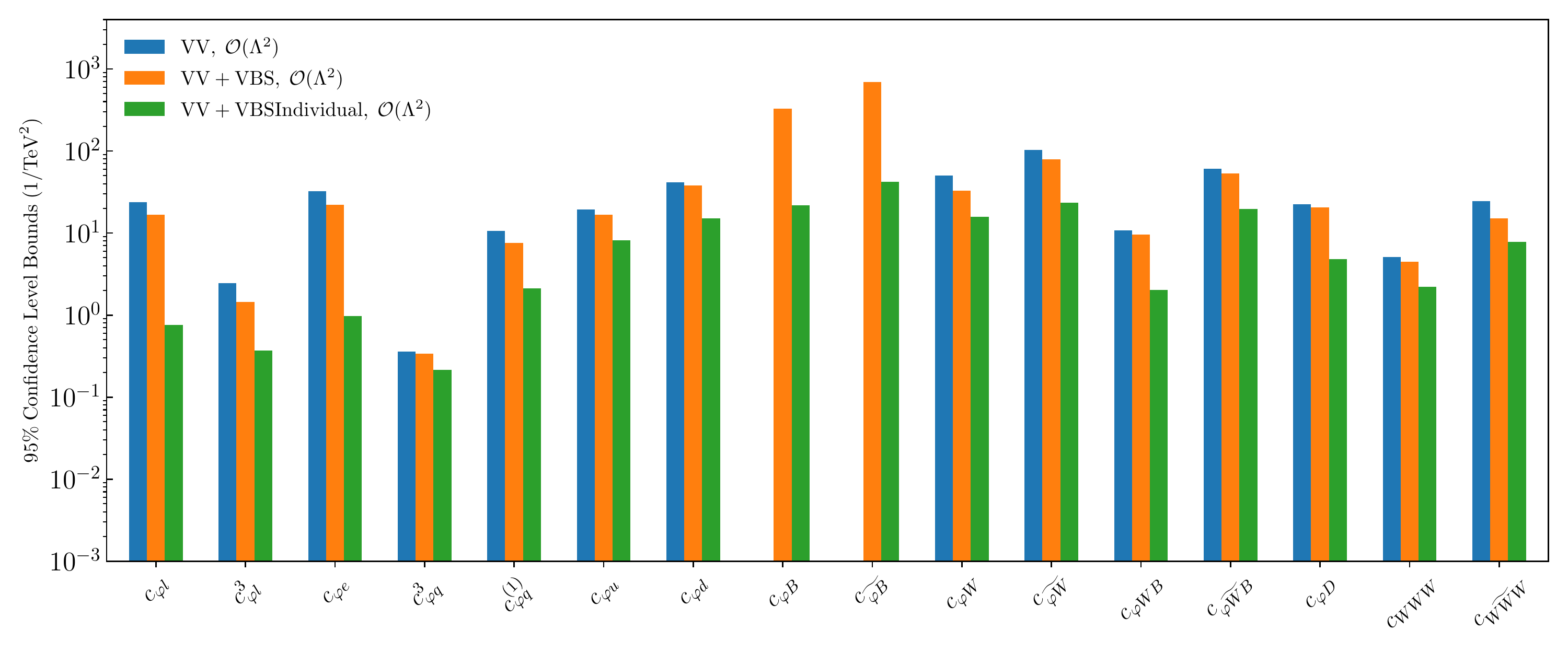} \\ %\hfill 
    \includegraphics[width=.95\textwidth]{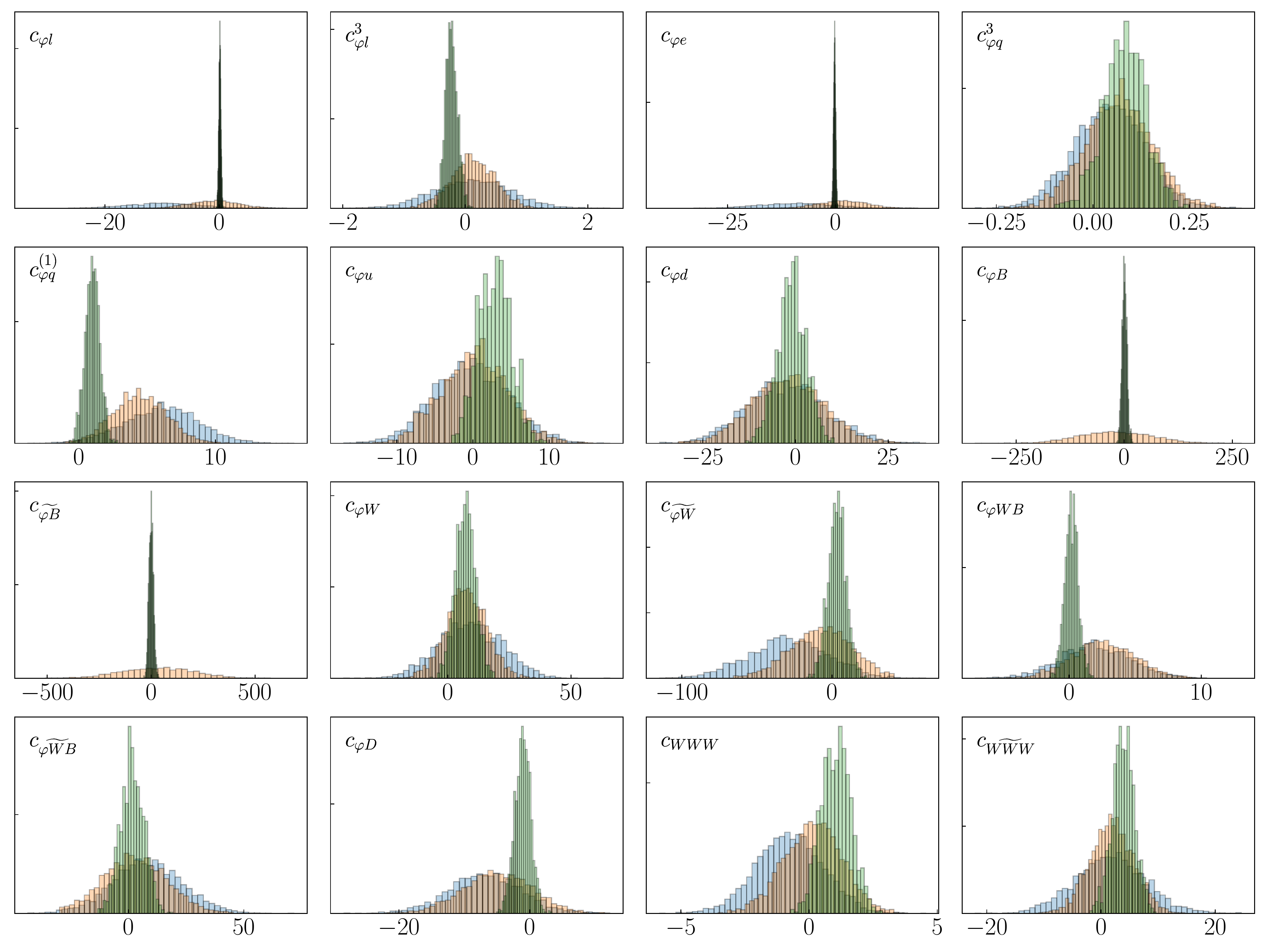}%\hfill
    \caption{Comparison of the magnitude of $95\%$ CL intervals (top) and posterior distribution samples (bottom) in three different settings: the VV only, the combined VV plus VBS marginalised and individual fits, at the linear level in the EFT expansion. The operators are in the Warsaw basis~\cite{Ethier:2021ydt}}\label{fig:main_results}
    
\end{center}
\end{figure}

In arXiv:2101.03180~\cite{Ethier:2021ydt},  we presented a fit of the LHC Run-2 electroweak sector, including the relevant operators of the dimension 6 SMEFT basis at the linear level. We found that VBS data, albeit still affected by relatively large experimental uncertainties due to the limited statistics, can still show an improvement with respect to the diboson dataset only. It has been shown in previous works that the main constraints to LHC fits come from Higgs data, and that diboson measurements have a relatively low impact on the final results, as shown in \cite{Ethier:2021bye,Ellis:2020unq}. Nevertheless, the inclusion of diboson and VBS data in this context represents a test for the electroweak symmetry breaking mechanism (EWSB) and the results show that the former is in very good shape. In figure~\ref{fig:main_results} we show a summary of our results where we observe that adding the VBS information helps to improve the precision of the diboson fit. This fact hints that, once  more precise VBS results are available (in particular, more differential data), the VBS can improve the VV fit substantially.

%%%%%%%%%%%%%%%%%%%%%%%%%%%%%%%%%%%%%%%%%%%%%%%%%%%%%%%%%%%%%%%%%%%%%%%%%%%%%%%%%%%%%%%%

\section{Two-dimensional fits}

\begin{figure}[!ht]
\begin{center}

    \includegraphics[width=.35\textwidth]{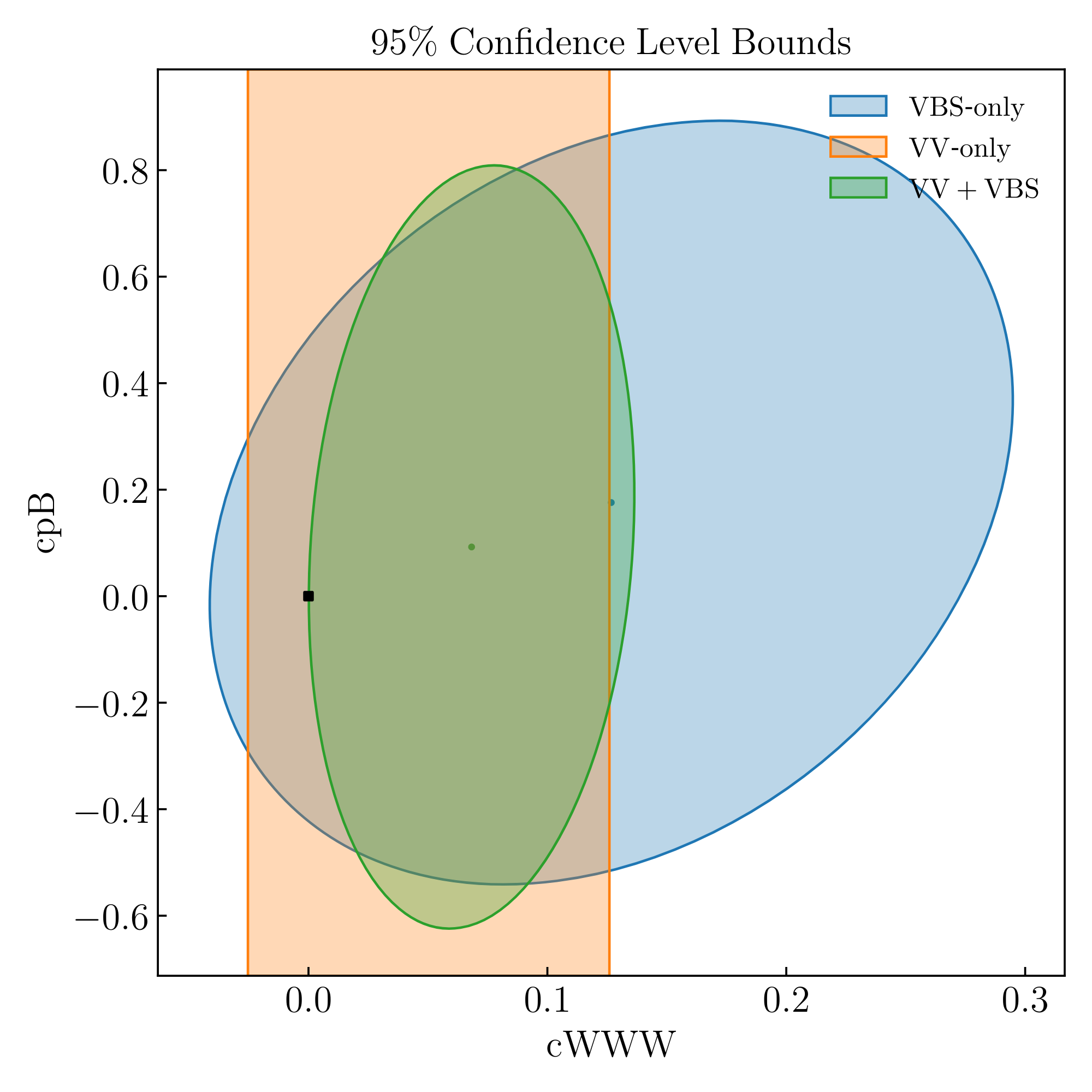}%\hfill
    \includegraphics[width=.35\textwidth]{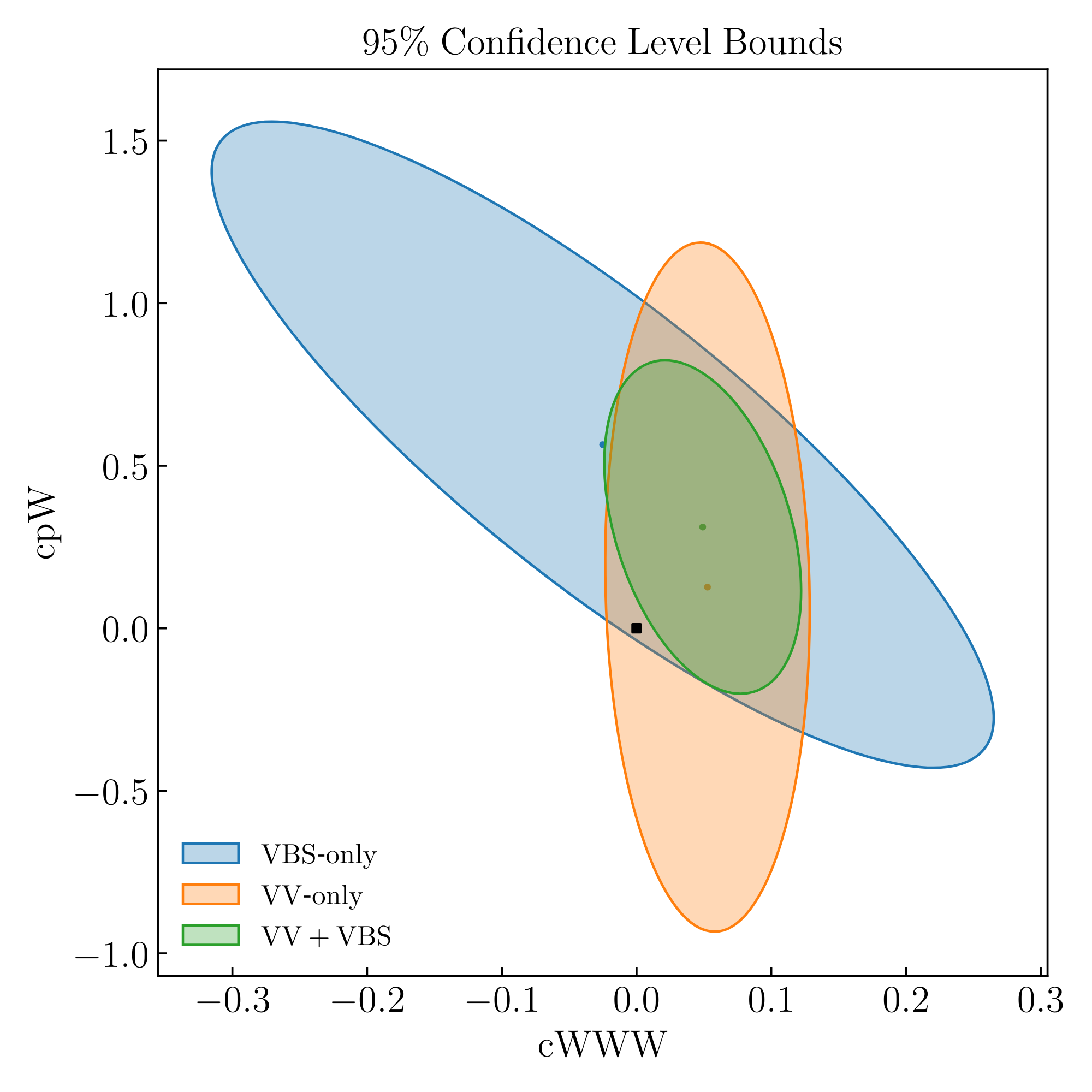}%\hfill
    \\[\smallskipamount]
    \includegraphics[width=.35\textwidth]{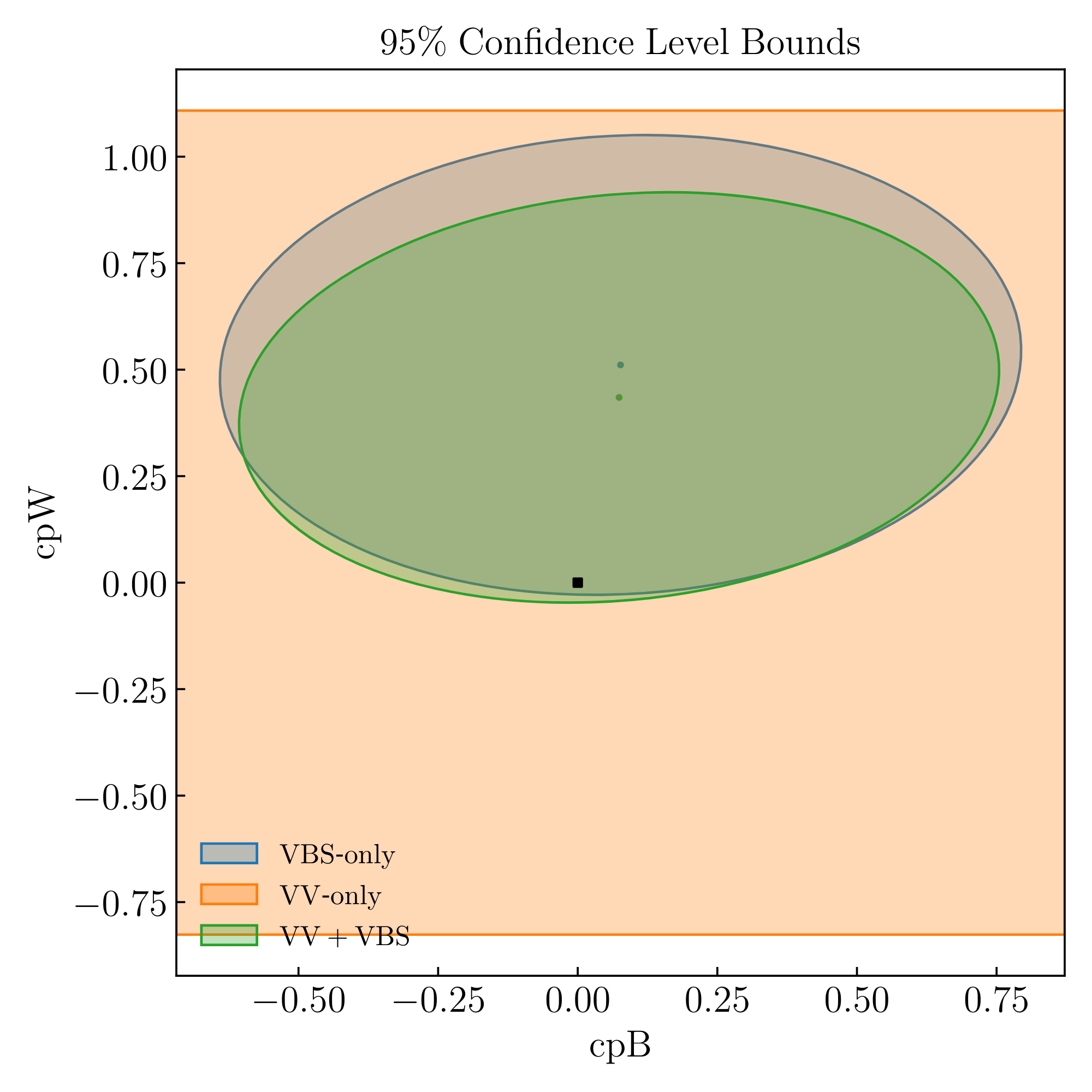}%\hfill
    \includegraphics[width=.35\textwidth]{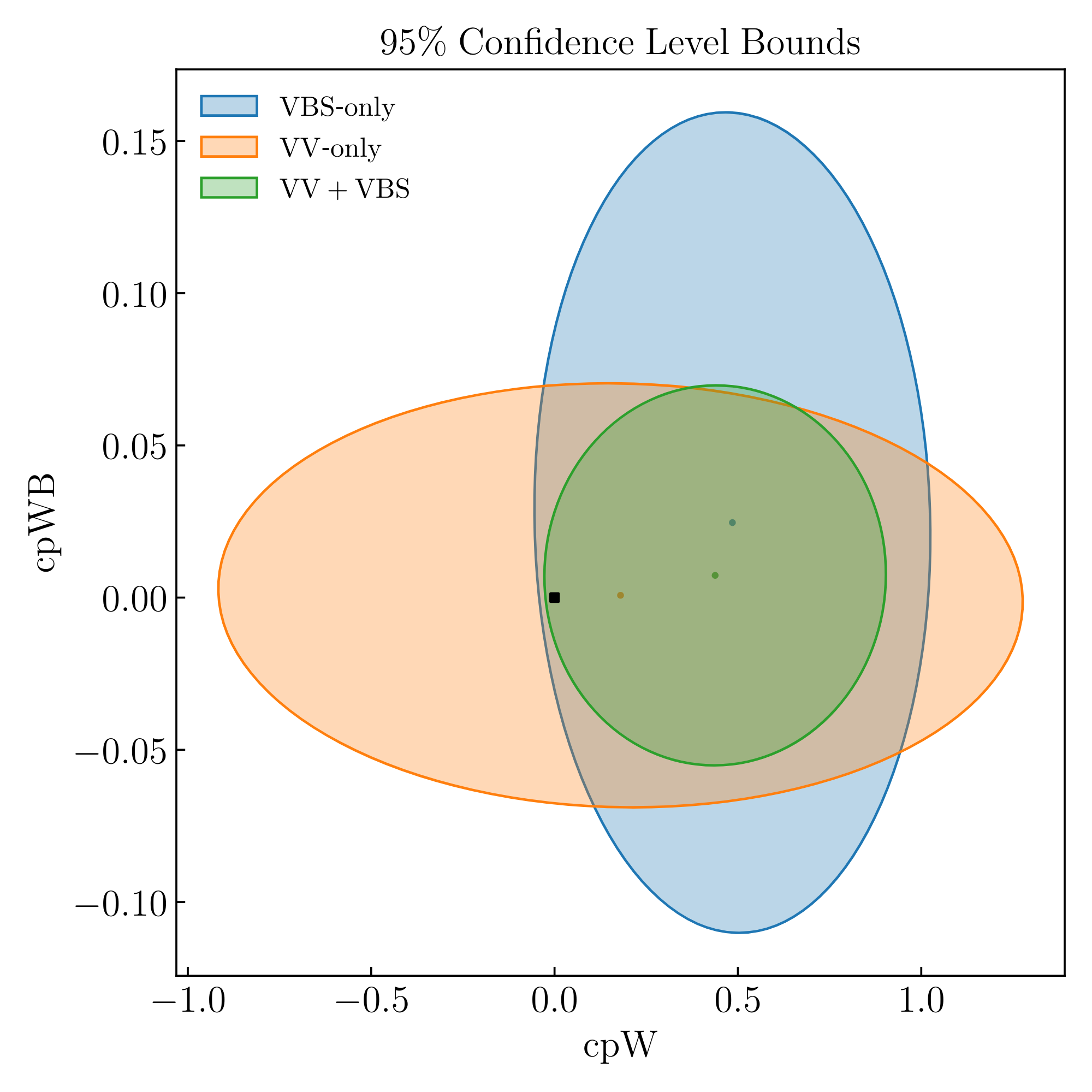}%\hfill
    \\[\smallskipamount]
    \caption{ We display the regions containing the $95\%$ of the posteriors distribution samples obtained for different data subsets and for the complete dataset (VV+VBS). Operator definitions are given in~\cite{Ethier:2021ydt}.}
    \label{fig:2dplots}
\end{center}
\end{figure}

In order to go one step further in our analysis, we have now incorporated a series of two-dimensional fits, in which only two operators are varied while all the rest are set to their SM values.
%This kind of fit is useful to understand correlations between different degrees of freedom
These are shown in figure~\ref{fig:2dplots}, where we appreciate the complementarity between the diboson and VBS channels. In the top row we show results for the $\mathcal{O}_{WWW}$ operator, the pure contribution to gauge couplings. VBS naturally complements diboson processes in this case as it contains the interplay between triple and quartic gauge couplings. If something was to be \emph{different} with the electroweak symmetry breaking mechanism, it would be likely to show up here, for example if the Higgs would not be an $SU(2)$ doublet, the ratio between these couplings would not necessarily be proportional to the Higgs vacuum expectation value and tensions would appear in this particular degree of freedom. The $\mathcal{O}_{\varphi W}$ and $\mathcal{O}_{\varphi B}$ operators, enter always by means of an $HVV$ coupling. In the bottom row we show results for these two: in the case of diboson processes, such operators contribute as contamination in the s-channel whereas in VBS they enter through Higgs boson exchanges in the t-channel, helping to remove their degeneracy. The last example shows the case of the $\mathcal{O}_{\varphi WB}$ operator, which accounts for the mixing in the neutral sector. This operator is deemed to be very well understood and constrained from electroweak precision measurements, but it is is still very interesting to see the complementarity between diboson and VBS in this case. 

These plots show the region containing the $95 \%$ of the posterior distribution samples obtained after the $\chi^2$ minimization. Whenever one degree of freedom is not affecting a specific process this region appears as band, while when both the two operators are constrained the CL bounds appear as an ellipse. 
In a two dimensional model at liner level in the EFT expansion is always possible to find out a linear combination of Wilson coefficients that cancels exactly the EFT contribution. This sort of flat direction coincide with the major axes of the ellipses presented here.

%%%%%%%%%%%%%%%%%%%%%%%%%%%%%%%%%%%%%%%%%%%%%%%%%%%%%%%%%%%%%%%%%%%%%%%%%%%%%%%%%%%%%%%%

\section{Moving to dimension 8 and computational bottlenecks}
Now that the global fits with the full dimension 6 basis at linear and quadratic level are well underway, it is time to start thinking on adding dimension 8 operators~\cite{Passarino:2019yjx}:
%
%\vspace{-0.3cm}
\begin{equation}
    \sigma_{EFT} = \sigma_{SM} + \frac{1}{\Lambda^2} \sigma_{int,6} + 
    \frac{1}{\Lambda^4} ( \sigma_{sq,6} + \sigma_{int,8} ) + \dots
\end{equation}
Many options are available in the market, including the full dimension 8 basis \cite{Murphy:2020rsh, Li:2020gnx}, as well as some works selecting a subset of process dependent operators. These can be used to estimate the effects of the dimension 8 contributions in particular channels, as shown in refs.\cite{Dawson:2021xei,Hays:2018zze}.
For the case of VBS processes, 
an ad-hoc basis~\cite{Eboli:2006wa} was defined to account for anomalous quartic gauge couplings (aQGCs), and it has been frequently adopted by the experimental collaborations. Several analyses including bounds on dimension 8 operators\footnote{These analyses include only this restricted set of dimension 8 operators, and no dimension 6 contribution.} have been released by the experimental collaborations in recent years. To name a few: \cite{CMS:2017zmo,ATLAS:2021pdg,CMS:2021gme,ATLAS:2020nlt}.

Adopting this approach has the drawback that the basis is not complete as it neglects by definition any contribution affecting triple gauge couplings: for example operators of the form $\mathcal{O}_8 \sim (H^{\dagger} H) \epsilon_{ijk} W_{\mu \nu}^i W_{\nu \rho}^j W_{\rho \mu}^k \, $ are not included.

To address this issue and given that this basis has now been widely used and experimental constraints are available, we suggest a three-step procedure:
\begin{enumerate}[itemsep=2pt,parsep=2pt]
    \item Perform a fit for the VBS channels including the dimension 6 basis as well as the aQCG one.
    \item Extract the relevant dimension 8 operators from the complete basis and repeat the procedure.
    \item Compare results and draw conclusions based on their consistency. 
\end{enumerate}
%
%%%%%%%%%%%%%%%%%%%%%%%%%%%%%%%%%%%%%%%%%%%%
%
Furthermore including dimension 8 terms from the aQGC basis, could be very interesting, since it allows for interactions that are traditionally forbidden by gauge invariance in the SM, for example vertices of the form $Z_{\mu}Z_{\nu}Z_{\rho}Z_{\sigma}$ or $Z_{\mu}Z_{\nu} \gamma_{\rho} \gamma_{\sigma}$.
 Non-SM-like couplings are important to ensure that we are not missing a possible trace of new physics. The available experimental bounds show that neutral aTCGs are very unlikely, but results on aQCGs are still not completely ruling them out. 
 % \hl{We are still investigating weather this approach could be consistent, since fully neutral aTCG will remain forbidden, in agreement with present experimental bounds.}

Additionally, it is important to consider different scenarios for the EWSB mechanism, which might shed light on the shape of the Higgs potential. 
For example, in the HEFT Lagrangian~\eqref{eq:HEFTlag}, a non-linear representation for the Higgs field is adopted. The physical $h$ field is deemed to be independent from the goldstone bosons $(w^+,w^-)$, for example by arranging them as a phase of the covariant derivative.
The EFT expansion then, is not performed by adding higher dimensional operators regularized but a cut off scale. Instead, the expansion is in terms of $h/v$, leading to a different power counting:

%The HEFT Lagrangian will then have the form:

\begin{equation}
    \mathcal{L}_{HEFT} =  \left(
    1 + a \frac{h}{v} + b \frac{h^2}{v^2} + \dots
    \right) \partial_{\mu} w^+ \partial^{\mu} w^- + 
    \frac{1}{2} \partial_{\mu} h \partial^{\mu} h + \dots
    \label{eq:HEFTlag}
\end{equation}
 This model has been studied in the particular context of VBS, the paradigmatic example of unitary violations, in different publications,\cite{Dobado:2017lwg,Delgado:2017cls,Delgado:2019ucx}. However extending it to reproduce the full LHC setup still represents some technical challenge: most Monte Carlo generators assume BSM models to be an extension of the SM, whereas in this Lagrangian the $SU(2)xU(1)$ structure is removed and reproducing experimental results becomes non-trivial. Specifically for a channel like VBS with a 6-fermion final state. 
 %Moreover the fermion structure in this Lagrangian is non-trivial, and reproducing such a 6-fermion final state, in the experimental baseline, represents a technical challenge.
 %
 This particular theory would particularly benefit from the Pseudo Observable approach that we discuss in the following section. 

%%%%%%%%%%%%%%%%%%%%%%%%%%%%%%%%%%%%%%%%%%%%%%%%%%%%%%%%%%%%%%%%%%%%%%%%%%%%%%%%%%%%%%

\section{Computational Bottlenecks and Pseudo Observables}

At this stage it is important to reflect on the computational demands of performing such a global fit.
The fitting tool used here, SMEFiT \cite{Ethier:2021bye}
is based on a sampling procedure, and can perform a quadratic fit with
$\mathcal{O}(50)$ operators in a a couple hours working on multiple CPU. The bottleneck though is on the prediction side. Reproducing accurately the experimental definitions, parton shower set-up and selection cuts, which are different for each collaborations, is a non-trivial exercise. To put it quantitatively: reproducing the SM prediction for a certain diboson process takes a couple 
of CPU hours for the generation and showering of $\mathcal{O}(500k)$ events, plus a few minutes for the analysis of the output files with {\tt{Rivet}}\footnote{This is the most accurate approach as experimental data are published in {\tt{HepData}} in the native {\tt{Rivet}} format ({\tt{.yoda)}}.}. The same procedure for a VBS process (6-particle final state) takes a few days already at LO.  

\begin{figure}[!ht]
\begin{center}
    \includegraphics[width=.5\textwidth]{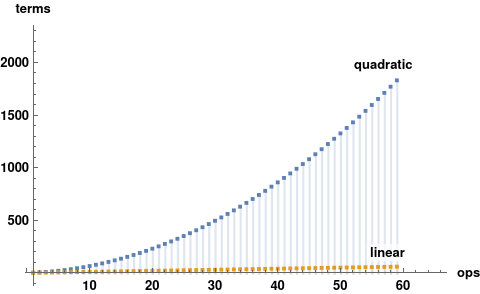}%\hfill
    \includegraphics[width=.5\textwidth]{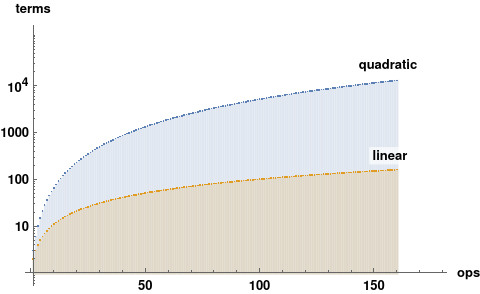}%\hfill
    \caption{Number of terms in the EFT expansion for a given observable. On the left we see the tendency up to 59 operators, the dimension 6 case assuming flavour universality. On the right we see how the numbers would grow as we add higher dimensional contributions. }\label{fig:nterms}
\end{center}
\end{figure}

Once the SM theory prediction agrees with the experimental one, one can move onto generating the different EFT predictions, for the case of $n$ operators, one needs to scale the CPU time (and disk storage) by $n$ for a linear prediction. Moving to quadratic terms, the situation worsens dramatically: we need to calculate additional $n$ predictions for each operator interfering with itself, as well as $\binom{n}{2}$ cross-terms. This tendency is illustrated in Figure~\ref{fig:nterms}. 

The main alternative to circumvent this issue would be to define new quantities which are easier to reproduce from the theory side, namely certain Pseudo Observables~\cite{Passarino:2010qk,Greljo:2015sla,Gainer:2018qjm}. Some works have pointed in this direction recently, with the main issues being that such observables are process-dependent (one needs to find the \emph{optimal observable} for a given process), and that, if a certain technique is used to process the data (typically a neural network), the details of the former would have to be passed on to the theory community alongside with the results. This issue has been discussed in the past given that published data for differential distributions at LHC often rely on machine learning techniques, and it is difficult to estimate their impact on the exact final event selection. This question has been highlighted again recently in~\cite{Arratia:2021otl,Cranmer:2021urp}, where some solutions where proposed. 
Another interesting approach is that of Matrix Element methods, a type of multivariate analysis techniques where predictions are compared at the level of the $S$-matrix elements, independent of phase space definitions. A recent interesting proposal in this direction is that of~\cite{Butter:2021rvz}.

Last but not least, there is the issue of \emph{precision}. It would be important to revisit our understanding of precision measurements and the proton structure, in the context of the EFT interpretations, to ensure we are not overlooking some new physics effects. Some works including the SMEFT effects in PDF fits have recently been presented in~\cite{Greljo:2021kvv}.
On this note, it would be also interesting to perform combined studies to understand the impact of the uncertainties that can be associated to the determination of the SM parameters, in the context of non-SM-like scenarios like HEFT.

\section{Conclusions}

Now that global fits of the dimension 6 basis in the LHC dataset are well underway, it is time to look into the near future and (re)design our strategy for the analysis of the forthcoming runs. The main challenges we find are two: 
On the more technical level, the inclusion of higher orders in the EFT expansion, needs of large computational resources. This issue might be circumvented in the future if the theory and experimental communities agree on a common framework for the publication of data and statistical models. 
On the broader side of the question, we have to ensure that our results don't fall into a \emph{self fulfilled prophecy}. By assuming Standard-Model-like structures we might be missing information on strongly coupled new physics.
Understanding the shape of the Higgs potential through direct $HH$ and $HHH$ production measurements seems to be slightly out of reach given the HL-LHC projections and alternative approaches should also be considered.
This task should be a priority for the collider phenomenology community, and EFT searches in the EW sector should illustrate this fact.

\section{Acknowledgements}

\noindent
We are very thankful to our collaborators J. J. Ethier and J. Rojo.


\begin{thebibliography}{99}

\bibitem{Ellis:2020unq}
J.~Ellis, M.~Madigan, K.~Mimasu, V.~Sanz and T.~You,
\emph{Top, Higgs, Diboson and Electroweak Fit to the Standard Model Effective Field Theory},
JHEP \textbf{04} (2021), 279
doi:10.1007/JHEP04(2021)279
 \href{https://arxiv.org/abs/2101.03180}{\tt{[arXiv:2012.02779]}}

\bibitem{Ethier:2021bye}
J.~J.~Ethier, G.~Magni, F.~Maltoni, L.~Mantani, E.~R.~Nocera, J.~Rojo, E.~Slade, E.~Vryonidou and C.~Zhang,
\emph{Combined SMEFT interpretation of Higgs, diboson, and top quark data from the LHC},
\href{https://arxiv.org/abs/2101.03180}{\tt{[arXiv:2105.00006]}}

\bibitem{Ethier:2021ydt}
    J. J. Ethier, R. Gomez-Ambrosio, G. Magni, and J. Rojo. 
    \emph{SMEFT analysis of vector boson scattering and diboson data from the LHC Run II}
     Eur. Phys. J. C Vol.81 N.6 (560) 2021,
   \href{https://arxiv.org/abs/2101.03180}{\tt{[arXiv:2101.03180]}}

\bibitem{Passarino:2019yjx}
G.~Passarino,
\emph{XEFT, the challenging path up the hill: dim = 6 and dim = 8}
\href{https://arxiv.org/abs/1901.04177}{\tt[arXiv:1901.04177]}

\bibitem{Murphy:2020rsh}
C.~W.~Murphy,
\emph{Dimension-8 operators in the Standard Model Eective Field Theory}
JHEP 10 (2020), 174,
\href{https://arxiv.org/abs/2005.00059}{\tt{[arXiv:2005.00059]}}

\bibitem{Li:2020gnx}
H.~L.~Li, Z.~Ren, J.~Shu, M.~L.~Xiao, J.~H.~Yu and Y.~H.~Zheng,
\emph{Complete set of dimension-eight operators in the standard model effective field theory}
Phys. Rev. D \textbf{104} (2021) no.1, 015026,
\href{https://arxiv.org/abs/2005.00008}{\tt{[arXiv:2005.00008]}}

\bibitem{Dawson:2021xei}
S.~Dawson, S.~Homiller and M.~Sullivan,
\emph{The Impact of Dimension-8 SMEFT Contributions: A Case Study}
\href{https://arxiv.org/abs/2110.06929}{\tt{[arXiv:2110.06929]}}

\bibitem{Hays:2018zze}
C.~Hays, A.~Martin, V.~Sanz and J.~Setford,
\emph{On the impact of dimension-eight SMEFT operators on Higgs measurements}
JHEP 02 (2019), 123
\href{https://arxiv.org/abs/2110.06929}{\tt{[arXiv:1808.00442]}}


\bibitem{Eboli:2006wa}
O.~J.~P.~Eboli, M.~C.~Gonzalez-Garcia and J.~K.~Mizukoshi,
\emph{p p ---\ensuremath{>} j j e+- mu+- nu nu and j j e+- mu-+ nu nu at O( alpha(em)**6) and O(alpha(em)**4 alpha(s)**2) for the study of the quartic electroweak gauge boson vertex at CERN LHC}
Phys. Rev. D 74 (2006), 073005
\href{https://arxiv.org/abs/hep-ph/0606118}{\tt{[arXiv:hep-ph/0606118]}}


\bibitem{CMS:2017zmo}
A.~M.~Sirunyan \textit{et al.} [CMS],
\emph{Measurement of vector boson scattering and constraints on anomalous quartic couplings from events with four leptons and two jets in proton\textendash{}proton collisions at $\sqrt{s}=$ 13 TeV}
Phys. Lett. B 774 (2017), 682-705,
\href{https://arxiv.org/abs/1708.02812}{\tt{[arXiv:1708.02812]}}

\bibitem{ATLAS:2021pdg}
G.~Aad \textit{et al.} [ATLAS],
\emph{Observation of electroweak production of two jets in association with an isolated photon and missing transverse momentum, and search for a Higgs boson decaying into invisible particles at 13 TeV with the ATLAS detector}
\href{https://arxiv.org/abs/2109.00925}{\tt{[arXiv:2109.00925]}}.

\bibitem{CMS:2021gme}
A.~Tumasyan \textit{et al.} [CMS],
\emph{Measurement of the electroweak production of Z$\gamma$ and two jets in proton-proton collisions at $\sqrt{s} =$ 13 TeV and constraints on anomalous quartic gauge couplings}
Phys. Rev. D 104 (2021), 072001,
\href{https://arxiv.org/abs/2106.11082}{\tt{[arXiv:2106.11082]}}

\bibitem{ATLAS:2020nlt}
G.~Aad \textit{et al.} [ATLAS],
\emph{Observation of electroweak production of two jets and a $Z$-boson pair with the ATLAS detector at the LHC}
\href{https://arxiv.org/abs/2004.10612}{\tt{[arXiv:2004.10612]}}


\bibitem{Dobado:2017lwg}
A.~Dobado, F.~J.~Llanes-Estrada and J.~J.~Sanz-Cillero,
\emph{Resonant production of Wh and Zh at the LHC}
JHEP 03 (2018), 159
\href{https://arxiv.org/abs/1711.10310}{\tt{[arXiv:1711.10310]}}

\bibitem{Delgado:2017cls}
R.~L.~Delgado, A.~Dobado, D.~Espriu, C.~Garcia-Garcia, M.~J.~Herrero, X.~Marcano and J.~J.~Sanz-Cillero,
\emph{Production of vector resonances at the LHC via WZ-scattering: a unitarized EChL analysis}
JHEP 11 (2017), 098
\href{https://arxiv.org/abs/1707.04580}{\tt{[arXiv:1707.04580]}}

\bibitem{Delgado:2019ucx}
R.~L.~Delgado, C.~Garcia-Garcia and M.~J.~Herrero,
\emph{Dynamical vector resonances from the EChL in VBS at the LHC: the WW case}
JHEP 11 (2019), 065
\href{https://arxiv.org/abs/1907.11957}{\tt{[arXiv:1907.11957]}}

%\bibitem{LHCHiggsCrossSectionWorkingGroup:2016ypw}
%D.~de Florian \textit{et al.} [LHC Higgs Cross Section Working Group],
%\emph{Handbook of LHC Higgs Cross Sections: 4. Deciphering the Nature of the Higgs Sector}
%\href{https://arxiv.org/abs/1610.07922}{[arXiv:1610.07922]}

\bibitem{Passarino:2010qk}
G.~Passarino, C.~Sturm and S.~Uccirati,
\emph{Higgs Pseudo-Observables, Second Riemann Sheet and All That}
Nucl. Phys. B 834 (2010), 77-115
\href{https://arxiv.org/abs/1001.3360}{\tt{[arXiv:1001.3360]}}

\bibitem{Greljo:2015sla}
A.~Greljo, G.~Isidori, J.~M.~Lindert and D.~Marzocca,
\emph{Pseudo-observables in electroweak Higgs production}
Eur. Phys. J. C 76 (2016) no.3, 158
\href{https://arxiv.org/abs/1512.06135}{\tt{[arXiv:1512.06135]}}

\bibitem{Gainer:2018qjm}
J.~S.~Gainer, M.~Gonz\'alez-Alonso, A.~Greljo, S.~Isakovi\'c, G.~Isidori, A.~Korytov, J.~Lykken, D.~Marzocca, K.~T.~Matchev and P.~Milenovi\'c, \textit{et al.}
\emph{Adding pseudo-observables to the four-lepton experimentalist\textquoteright{}s toolbox}
JHEP 10 (2018), 073
\href{https://arxiv.org/abs/1808.00965}{\tt{[arXiv:1808.00965]}}


\bibitem{Arratia:2021otl}
M.~Arratia, A.~Butter, M.~Campanelli, V.~Croft, D.~Gillberg, K.~Lohwasser, B.~Malaescu, V.~Mikuni, B.~Nachman and J.~Rojo, \textit{et al.}
\emph{Presenting Unbinned Differential Cross Section Results}
\href{https://arxiv.org/abs/2109.13243}{\tt{[arXiv:2109.13243]}}.

\bibitem{Cranmer:2021urp}
K.~Cranmer, S.~Kraml, H.~B.~Prosper, P.~Bechtle, F.~U.~Bernlochner, I.~M.~Bloch, E.~Canonero, M.~Chrzaszcz, A.~Coccaro and J.~Conrad, \textit{et al.}
\emph{Publishing statistical models: Getting the most out of particle physics experiments}
\href{https://arxiv.org/abs/2109.04981}{\tt{[arXiv:2109.04981]}}

\bibitem{Butter:2021rvz}
A.~Butter, T.~Plehn and N.~Soybelman,
\emph{Back to the Formula -- LHC Edition}
\href{https://arxiv.org/abs/2109.10414}{\tt{[arXiv:2109.10414]}}

\bibitem{Greljo:2021kvv}
A.~Greljo, S.~Iranipour, Z.~Kassabov, M.~Madigan, J.~Moore, J.~Rojo, M.~Ubiali and C.~Voisey,
\emph{Parton distributions in the SMEFT from high-energy Drell-Yan tails}
JHEP 07 (2021), 122
\href{https://arxiv.org/abs/2104.02723}{\tt{[arXiv:2104.02723]}}

%\bibitem{Agrawal:2019bpm}
%P.~Agrawal, D.~Saha, L.~X.~Xu, J.~H.~Yu and C.~P.~Yuan,
%\emph{Determining the shape of the Higgs potential at future colliders}
%Phys. Rev. D \textbf{101} (2020) no.7, 075023
%doi:10.1103/PhysRevD.101.075023
%\href{https://arxiv.org/abs/1907.02078}{\tt{[arXiv:1907.02078]}}.

\end{thebibliography}
\end{document}